\documentclass[doublecol]{epl2}

\usepackage{graphicx}
\usepackage{color}

\title{A bottleneck model for bidirectional transport
controlled by fluctuations}

\author{Asja Jeli\'c\inst{1} \and C\'ecile Appert-Rolland\inst{1} \and Ludger Santen\inst{2}}
\institute{
  \inst{1} Laboratory of Theoretical Physics,
  CNRS (UMR 8627), University Paris-Sud\\
  B\^atiment 210, F-91405 ORSAY Cedex, France.\\
  \inst{2} Fachrichtung Theoretische Physik, Universit\"at des Saarlandes
  \\
  D-66123 Saarbr\"ucken, Germany.}
\shortauthor{A. Jeli\'c \etal}


\abstract{
We introduce a new model to study the oscillations of opposite
flows sharing a common bottleneck and moving on two Totally
Asymmetric Simple Exclusion Process (TASEP) lanes.
We provide a theoretical analysis of the phase diagram, valid
when the 
flow in the bottleneck
is dominated by local stationary states.
In particular, we predict and find an inhomogeneous high
density phase, with a striped spatio-temporal structure.
At the same time, our results also show that some other
features of the model cannot be explained by the stationarity
hypothesis and require consideration of the transients
in the bottleneck at each reversal of the flow.
In particular, we show that
for short bottlenecks, the capacity of the system is at least as
high as for uni-directional flow, in spite of having to
empty the bottleneck at each reversal
- a feature that can be explained only by efficient transients.
Looking at more sensitive quantities like the distribution of
flipping times, we show that, in most regimes, the bottleneck
is driven by rare fluctuations and descriptions beyond the
stationary state are required.}

\begin{document}

\maketitle

The understanding of the macroscopic behavior of
complex systems out of equilibrium is one of the main
challenges in modern statistical mechanics. A common
feature of many non-equilibrium systems is the presence
of a current in their stationary state, in contrast to
equilibrium.
A general framework describing these systems is still
lacking, though the importance of current large
deviations and their link with fluctuations theorems
has been emphasized \cite{ft}.

The absence of such
general framework motivated the study of many
oversimplified microscopical models, among which
exclusion processes have been reference systems because
they allow for extremely precise numerical results
and exact analytical solutions in some cases.
Exclusion
processes are simple models defined on a discrete --usually
one-dimensional-- lattice, on which particles hop from site
to site.
Their role as a reference system is of particular
importance for exact calculations of large deviations,
the non-equilibrium counterpart of the free energy
\cite{derrida0711,chou_m_z11}.
At the same time, exclusion processes are
a flexible tool to model various physical systems.
Indeed, they capture the correct
collective behavior of systems with very different length
scales from social individuals, such as pedestrians \cite{chowdhury_s_n05},
vehicles \cite{review_traffic}, ants \cite{prl_ants,chowdhury_s_n05}, to molecular systems as
molecular
motors \cite{chowdhury_s_n05,chou_m_z11}, and microscopical ones as quantum dots \cite{quantum_dot}. 
Furthermore exclusion processes are closely related \cite{asepvskpz} to
growth phenomena described by the celebrated Kardar-
Parisi-Zhang equation \cite{kpz}, 
and the universality of these
phenomena has been confirmed in recent experiments in
liquid crystals \cite{exper}. 

For the asymmetric simple exclusion process (ASEP) the full current probability
distribution has been calculated for different boundary conditions \cite{derrida93c,current}.
However, while the ASEP captures the proper behavior of particles moving 
on a single lane, it is obviously not appropriate to 
describe situations in which the particles (such as individuals or cars)
move in few intersecting lanes. 
To describe such systems several one-dimensional exclusion processes must be
coupled and the task of determining the current distributions becomes harder.
Models for interacting parallel lanes (to describe e.g.\ the traffic on highways)
have been introduced and studied 
\cite{reichenbach_f_f06}.
In particular, in the so-called bridge
models, two lanes share a finite number of sites
(the `bridge') \cite{evans95a}.
Such systems have attracted a large interest due to the
symmetry breaking that occurs in most cases
\cite{evans95a,bridge_models}.

\begin{figure}[t] 
  \centering
   \includegraphics[width=0.45\textwidth]{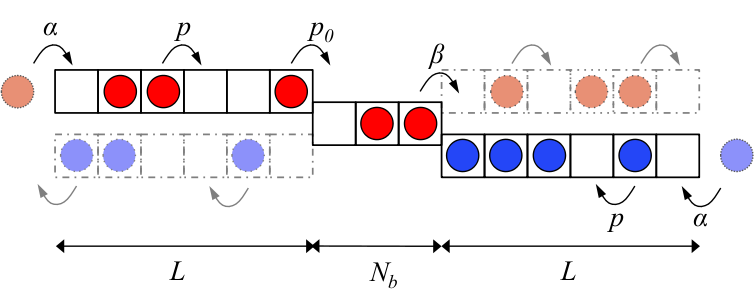}
   \caption{Schematic representation of the model.}
   \label{fig:model}
\end{figure}

In this letter, we introduce a new model which couples 
two totally asymmetric simple exclusion process (TASEP) lanes with oppositely directed flows
sharing a common bottleneck (see Fig. \ref{fig:model}).
In contrast to the bridge models,
exchange of oppositely moving particles is not possible inside the bottleneck,
because we impose that only particles going in one
direction can go through the bottleneck at a given time.
Therefore, particles going in the opposite direction
have to wait until the {\em whole} bottleneck is empty
before being allowed to go through.
Our model can be seen as a representation of
opposite pedestrian flows crossing e.g.\ at a door. 
The model is also relevant
for other systems such as multiphase flows, or
bidirectional molecular traffic across nuclear pores \cite{kapon08}.

We study this model with a combination of analytical and numerical techniques.
The property that makes this model qualitatively
different from previous ones 
is that the condition for reversing the flow inside the bottleneck
(i.e.\ for having an empty bottleneck) is a rare event as soon as the bottleneck exceeds a few sites.
Indeed,  
the dynamics of this system is
driven by rare fluctuations, and is non trivial.
For example, there exists a regime in which stop and go waves
invade the whole system.
Also, a counterintuitive feature of the model is that
it can sustain in the bottleneck a current 
{\em higher} than the maximal current of
a single lane system of infinite size.

\section{The model}
\label{sec:model}

We define now the model more precisely.
Particles 
move on two parallel tracks, modeled as two
TASEP
lanes.
Both lanes share the same bottleneck of length $N_b$.
We call `+' (resp.\ `-') the particles moving from left to
right (right to left), represented as red
(blue) particles in Fig. \ref{fig:model}.
We consider explicitly only the lanes of incoming particles (of length $L$).
Thus outgoing lanes are shadowed in Fig. \ref{fig:model}.
Random sequential update is considered.
Particles enter the lane with rate $\alpha$,
hop forward with rate $p$ if the next site
is empty, 
and
leave the bottleneck with rate $\beta$.
A particle enters the bottleneck with rate $p_0$ only
if (i) there is no particle of the other species inside
the bottleneck
and (ii) the first site of the bottleneck is empty.
We assume $p=p_0=1$,
leaving only $\alpha$ and $\beta$ as the model parameters, in addition to the bottleneck and lanes lengths $N_b$ and $L$.

\section{Phase diagram}
\label{sec:phase-diagram}

\begin{figure}[t]
\begin{center}
\includegraphics[width=0.4\textwidth]{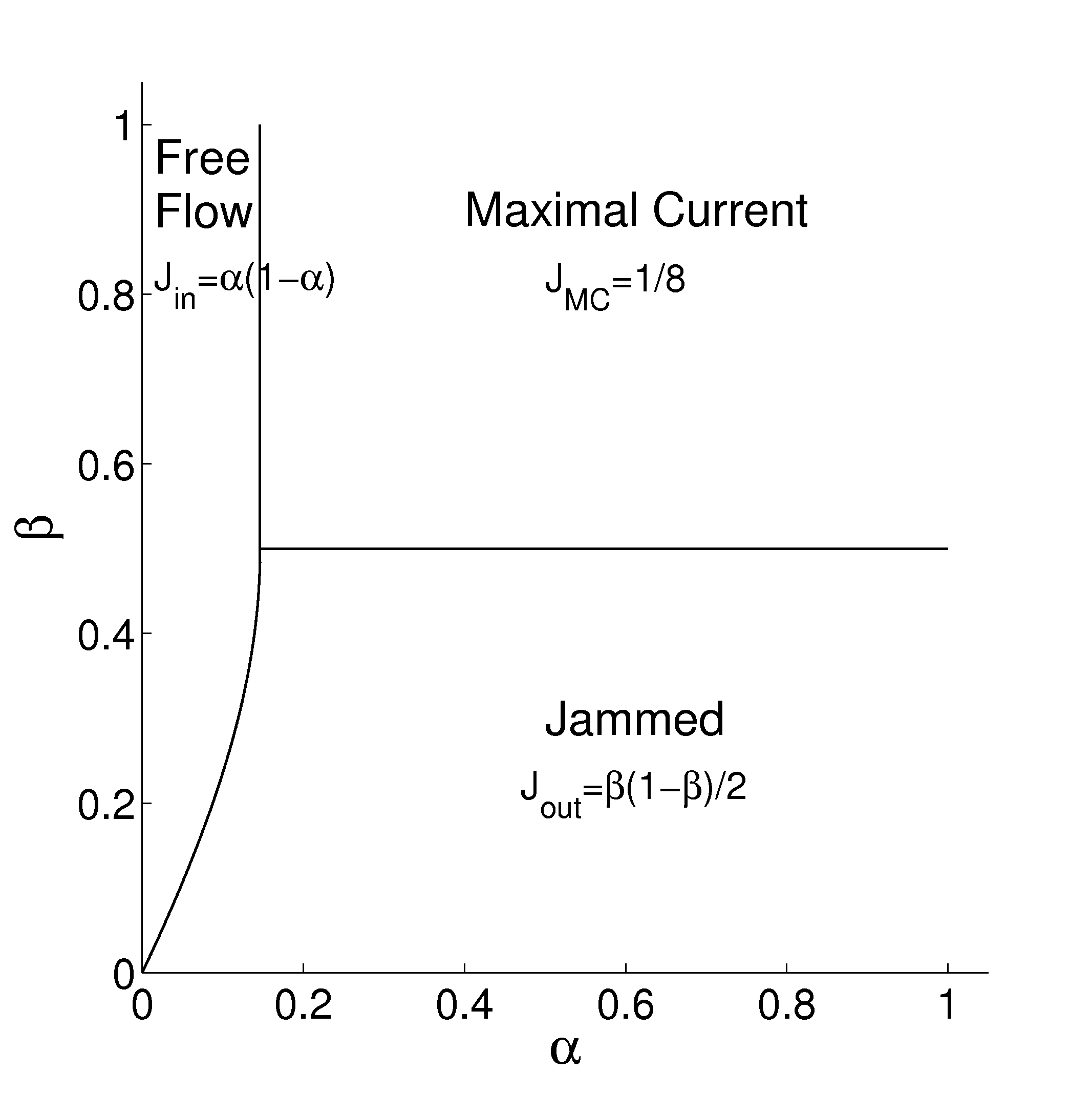}
\end{center}
\caption{Phase diagram of the bottleneck model,
predicted by the phenomenological approach.
}
\label{fig:phase-diagram-MF}
\end{figure}

By using the stationary properties of the TASEP,
we infer
the phase diagram (given in Fig. \ref{fig:phase-diagram-MF})
of the system with bottleneck
from simple phenomenological arguments. 
In the following, currents refer in general to the current
of one given species of particles (and not the total current).

For low $\alpha$ values, we expect a free flow (FF) phase
driven by the entrance rate.
For each species, the corresponding particle current and density are
\begin{equation}
J_{in}=\alpha(1-\alpha); \;\;\;\; \rho_{in} = \alpha.
\label{eq_jin}
\end{equation}
For large $\alpha$ 
and small $\beta$, we expect an exit driven
jammed (JJ) phase.
When a species goes through the
bottleneck for a long enough time, the system should
become equivalent to a one-lane TASEP in stationary
state, and the current should be $\beta (1-\beta)$.
As the use of the bottleneck is shared between the two
types of particles, each species can go through only
half of the time.
Neglecting the effects of the transients, on average
the current is 
\begin{equation}
J_{out}=\frac{1}{2} \beta(1-\beta).
\label{eq_jout}
\end{equation}
We shall see that the particle distribution
in incoming lanes is not homogeneous, and that, in contrast
to current, density cannot be obtained from this mean field like 
approach.

The FF/JJ boundary
is given by the solution of $J_{in} = J_{out}$, satisfying
$\alpha_c(\beta) \to 0$ when $\beta \to 0$, i.e.
\begin{equation}
\alpha_c (\beta)=\frac{1}{2}-\frac{1}{2}\sqrt{1-2\beta(1-\beta)}.
\label{boundary1}
\end{equation}
The FF phase corresponds to $\alpha < \alpha_c(\beta)$.

When $\alpha>1/2$ and $\beta>1/2$, the single lane
TASEP is in the maximal current (MC) phase, with a current
equal to $1/4$.
Again, if we neglect transients in the 
bottleneck, the
average current must be half of this value:
\begin{equation}
J_{MC}=1/8.
\label{eq_jmc1}
\end{equation}
The MC/FF boundary,
obtained from $J_{in} = J_{MC}$,
is given by
$
\alpha_c = (1-\sqrt{1/2})/2 \simeq 0.146,
$
while for the MC/JJ boundary,
$J_{out} = J_{MC}$ provides
$
\beta_c = \frac{1}{2}.
$
\begin{figure}[t]
\begin{center}
\includegraphics[width=0.423\textwidth]{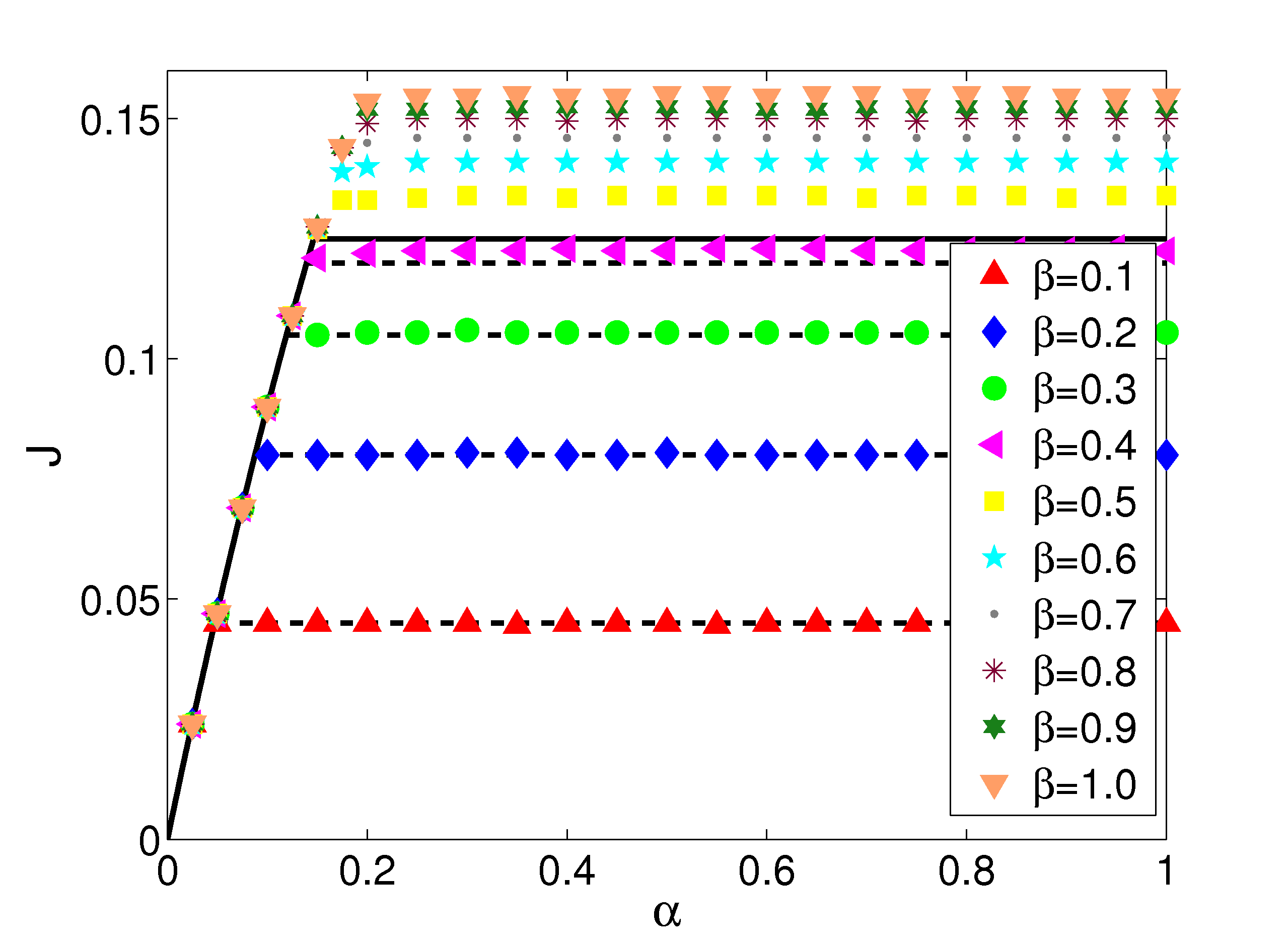}
\end{center}
\caption{
Average current as a function of $\alpha$, 
for different values of $\beta$. Symbols indicate 
numerical results obtained from Monte Carlo simulations
for $L = 50$ and $N_b = 4$. 
The solid line gives the theoretical predictions for $\beta > 1/2$, 
i.e.\ both for the free flow and the maximal current phase. 
Dashed lines give predictions for $\beta < 1/2$ for 
the jammed phase with $\alpha > \alpha_c (\beta)$.
}
\label{fig:current-betadiff}
\end{figure}

\begin{figure}[t]
\begin{center}
\includegraphics[width=0.40\textwidth]{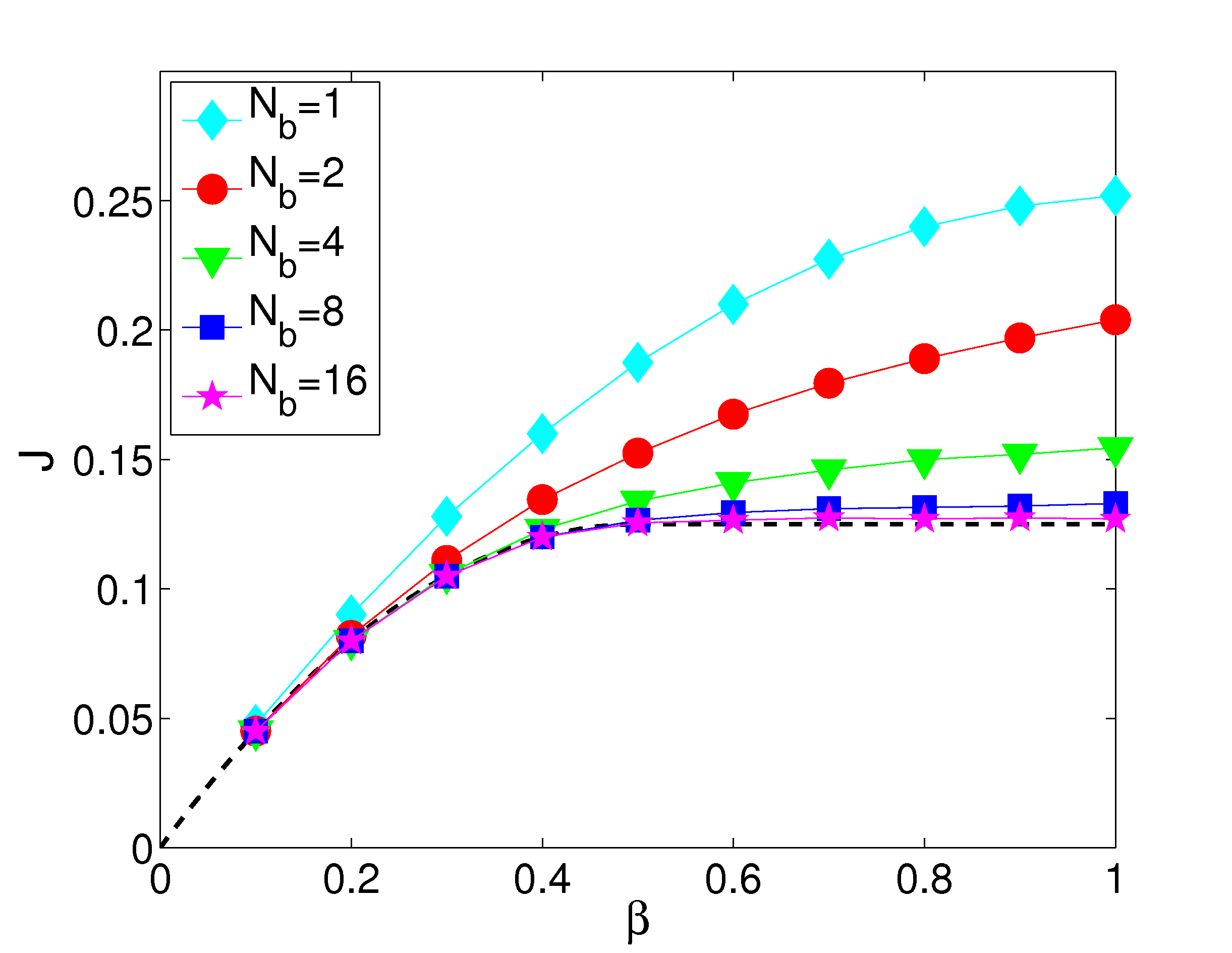}
\end{center}
\caption{Average current against
$\beta$
for different
$N_b$,
and $L=50$, $\alpha=0.6$.
The dashed line gives the theoretical predictions
for $\alpha=0.6$, both for the jammed and MC phase.
For large bottlenecks the current saturates to
$J_{MC}=1/8$.
}
\label{fig:current-nbdiff}
\end{figure}

Monte Carlo simulations confirm these
mean field predictions both for current and density in the FF phase
[Eq. (\ref{eq_jin})],
and for the current in the JJ phase [Eq. (\ref{eq_jout})]
for values of $\beta$ 
not close to the MC boundary 
(see Fig.\ \ref{fig:current-betadiff}).
By contrast, the current in the MC phase
is systematically underestimated by
this phenomenological
approach.
However, as seen in Fig. \ref{fig:current-nbdiff},
the maximal current value (\ref{eq_jmc1})
is recovered
for long bottlenecks, for which
a stationary state can
be established in the bottleneck.
For shorter bottlenecks, currents {\em higher} than
the stationary maximal current $1/4$ can be obtained.

Interestingly, with our bi-directional model,
we can even reach the capacity of a finite-size
system of size $N_b$ \cite{derrida93c} 
which would be fed by two incoming lanes,
each transporting the maximal current and injecting
particles on the same end of the bottleneck
(in this case, flow would be unidirectional in
the bottleneck and no flux reversal would be needed).

Hence the counterintuitive result
according to which it is at least as efficient
to have particles
coming from both ends of the bottleneck
than having all these particles coming from the same end.
This high capacity can be obtained in spite of the fact that
bidirectional traffic requires to empty the bottleneck at each
flux reversal, a limitation which is compensated by very
efficient transient states.
Indeed, at each reversal of the current in the bottleneck,
particles enter an empty system.
Their motion is not hindered by predecessors,
and thus high fluxes can be achieved.

\section{Spatio-temporal structures}
\label{sect_st}

To characterize the dynamics of the system,
we use a domain wall approach \cite{kolomeisky98,santen_a02}
describing the system at a mesoscopic scale.
We consider only the FF and JJ
phases, since the domain wall approach is not appropriate
to describe the MC phase (which 
has long-range correlations).
In this approach,
we neglect transients inside
the bottleneck, and assume that currents and densities
are given by stationary
expressions.

First we describe the FF phase, and
take the point of view of `+' particles.
When the bottleneck is closed, `+' particles
accumulate in front of the bottleneck,
forming a queue of density $1$.
The upstream end of the queue can be seen as a discontinuity
(or a wall) separating the queue
from the bulk.
A new wall is created when the bottleneck opens.
The queuing particles feed the bottleneck with
a high effective injection rate, and a high density domain 
is installed in the bottleneck.
Ignoring transients, this jammed domain
imposed by the exit has density $1-\beta$
and current $\beta(1-\beta)$.
Due to mass conservation, 
the wall between the jammed domain and the queue
moves backwards with velocity $1-\beta$ until
the whole queue is dissolved.
Then, at the separation between the bulk FF phase
and the jammed domain localized at the exit, a new wall forms
and moves
forward with velocity 
$
V = \frac{\left[\beta(1-\beta) - \alpha(1-\alpha)\right]}{(1-\alpha-\beta)}.
$
In order to
understand which phase invades the bulk, we have to
determine whether the FF domain will 
reach the entrance of the bottleneck {\em before}
it closes again.
This is the case if
\begin{equation}
\alpha (1-\alpha) \le \frac{1}{2} \beta(1-\beta),
\label{dw1}
\end{equation}
and assuming that the bottleneck is open and closed
for the same period of time $\tau$ (which is true on average).
Then the queue and the
jammed domain stay localized near the bottleneck,
and the bulk FF phase can be sustained.
If condition (\ref{dw1}) is not fulfilled, the
bulk FF phase is slowly invaded by the queue, 
and cannot survive over long times.
Note that
Eq. (\ref{dw1})
is indeed identical to condition (\ref{boundary1})
for the FF/JJ boundary.

\begin{figure}[t]
\begin{center}
\includegraphics[width=0.415\textwidth]{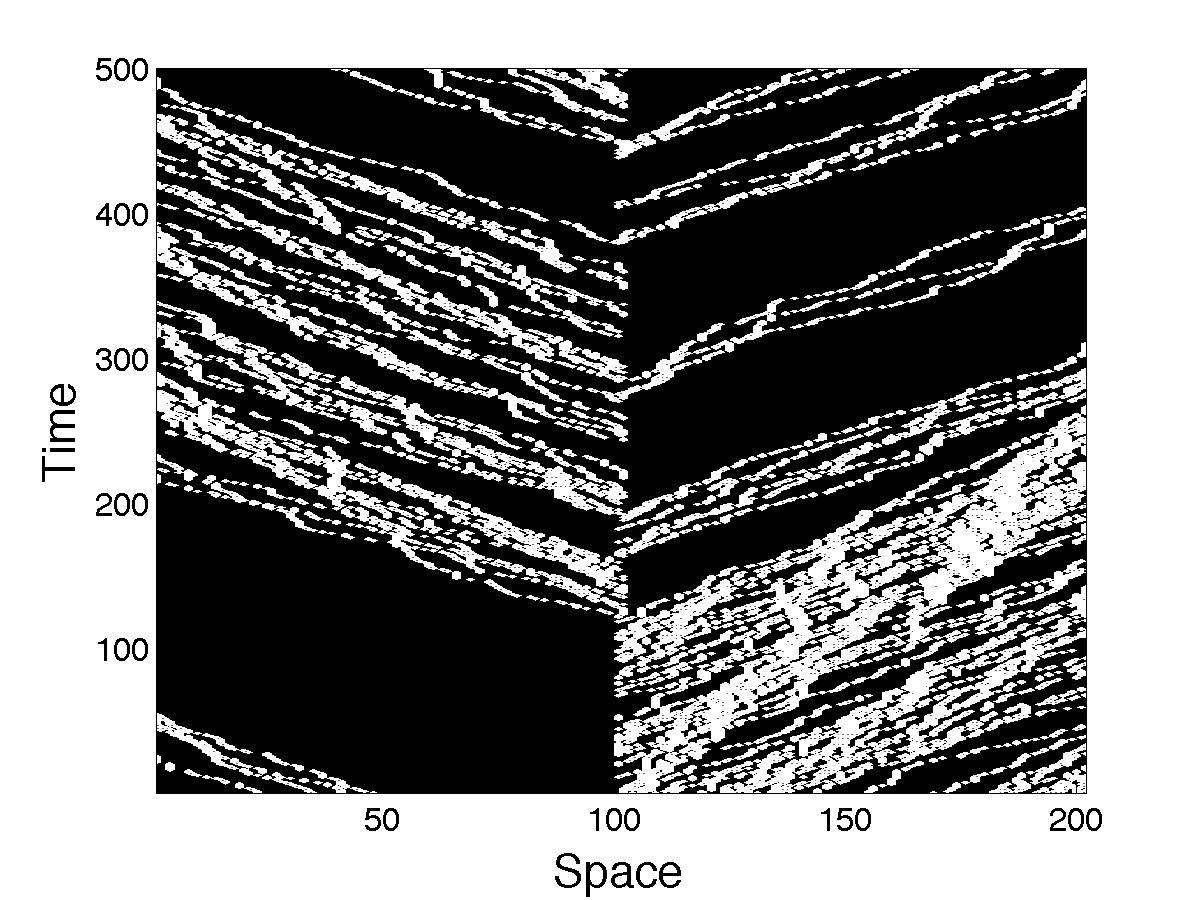}
\end{center}
\caption{
Spatio-temporal plot in the jammed
phase ($\alpha = 0.6$, $\beta = 0.2$, $L=100$, $N_b=2$).
Both incoming TASEP lanes
are shown, with the bottleneck in the middle.
Particles (empty sites) are represented as black (white) spots.
}
\label{fig:spacetime}
\end{figure}
Now we apply the same 
coarse-grained approach
to the JJ phase to show that a homogeneous
bulk density is not possible.
Indeed, $\rho_{bulk}$ should satisfy $\rho_{bulk}(1-\rho_{bulk}) = J_{out}$.
However, when the bottleneck is closed, a queue of density $1$ is
formed. The wall between this queue and the bulk
density moves backwards with velocity $\rho_{bulk}$.
Instead, when the bottleneck is open,
the wall separating 
the exit driven jammed domain and  the queue
moves backwards only with velocity $1-\beta$.
Thus, it cannot catch up with the previous wall,
and the queue can never be entirely dissolved.
When the bottleneck closes again, a new queue of density $1$
is formed, whose rear end also moves with velocity
$1-\beta$ inside the exit driven phase.
Then, regions of queuing particles
with density $1$ alternate with regions of 
density $1-\beta$ and current $\beta(1-\beta)$, 
resulting in an overall
striped jammed phase.
This is confirmed by 
spatio-temporal plot in Fig. \ref{fig:spacetime} obtained from simulations.
Averaging over the stripes gives an actual bulk density
\begin{equation}
\rho_{out} = 1-\frac{\beta}{2},
\label{eq_rhoout}
\end{equation}
also confirmed by the Monte Carlo simulations.

\section{Distribution of the oscillation periods}

\begin{figure}[t]
\begin{center}
\includegraphics[width=0.40\textwidth]{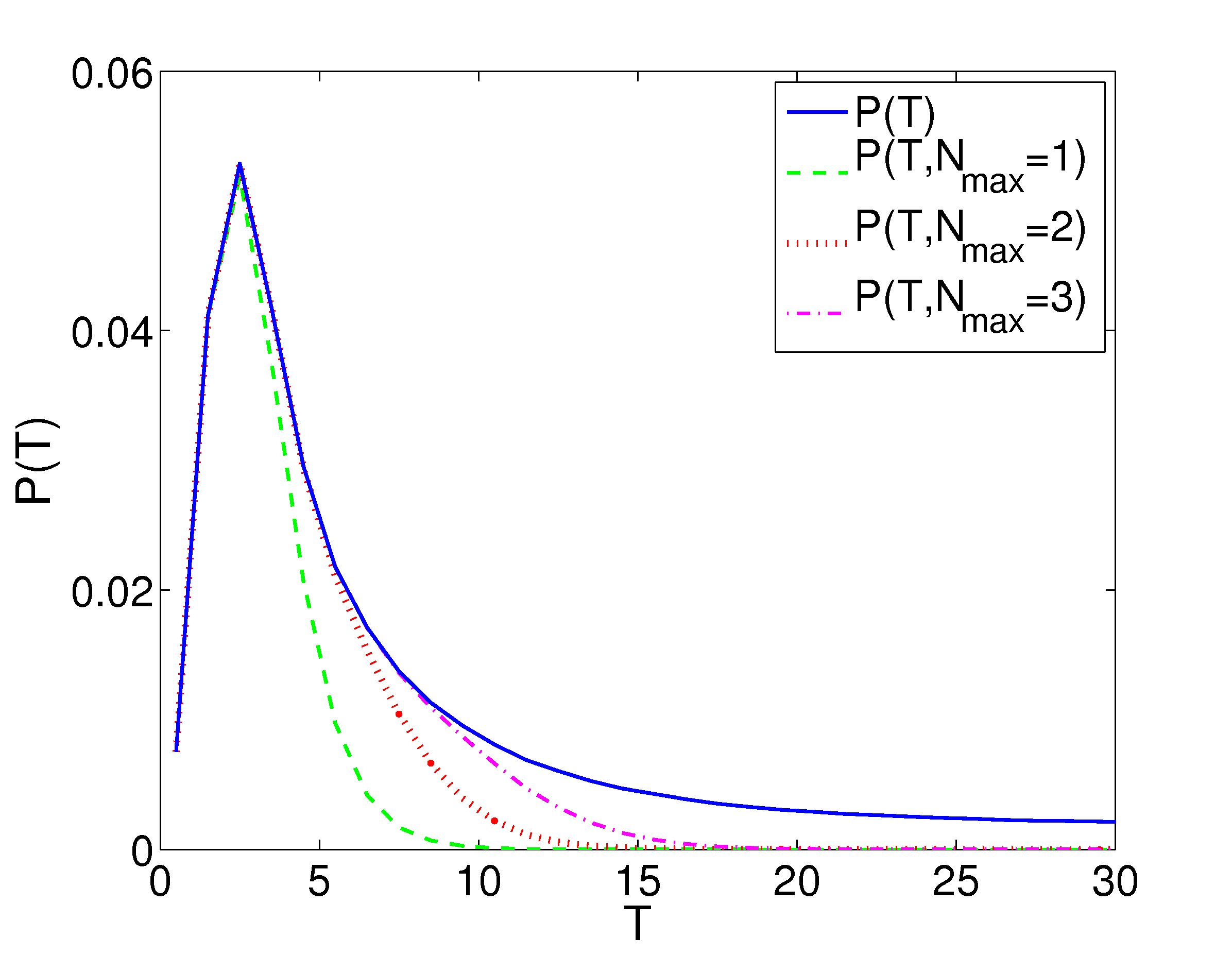}
\end{center}
\caption{
Distribution $P(T)$ 
for $N_b=4$, $\alpha = 0.6$, $\beta = 0.2$ and $L=400$.
The blue solid line is the whole distribution.
The other curves are the contributions to $P(T)$
corresponding to the passage of at most $N_{max}$
particles
through the bottleneck
during the time interval $T$.
}
\label{fig:distT}
\end{figure}

Until now, 
we considered the average value of
the oscillation
period $\tau$, which is a fluctuating variable.
To perform a quantitative analysis, we define $T$ as
the time during which the bottleneck
is occupied by at least one particle of
the type under consideration.
After $T$, bottleneck is empty and
a particle of either the same or opposite type
can enter.
Thus $T$ is not identical to $\tau$
but strongly related to it.
We now consider the probability distribution 
$P(T)$, shown in Fig. \ref{fig:distT}.

\begin{figure}[t]
\begin{center}
\includegraphics[width=0.425\textwidth]{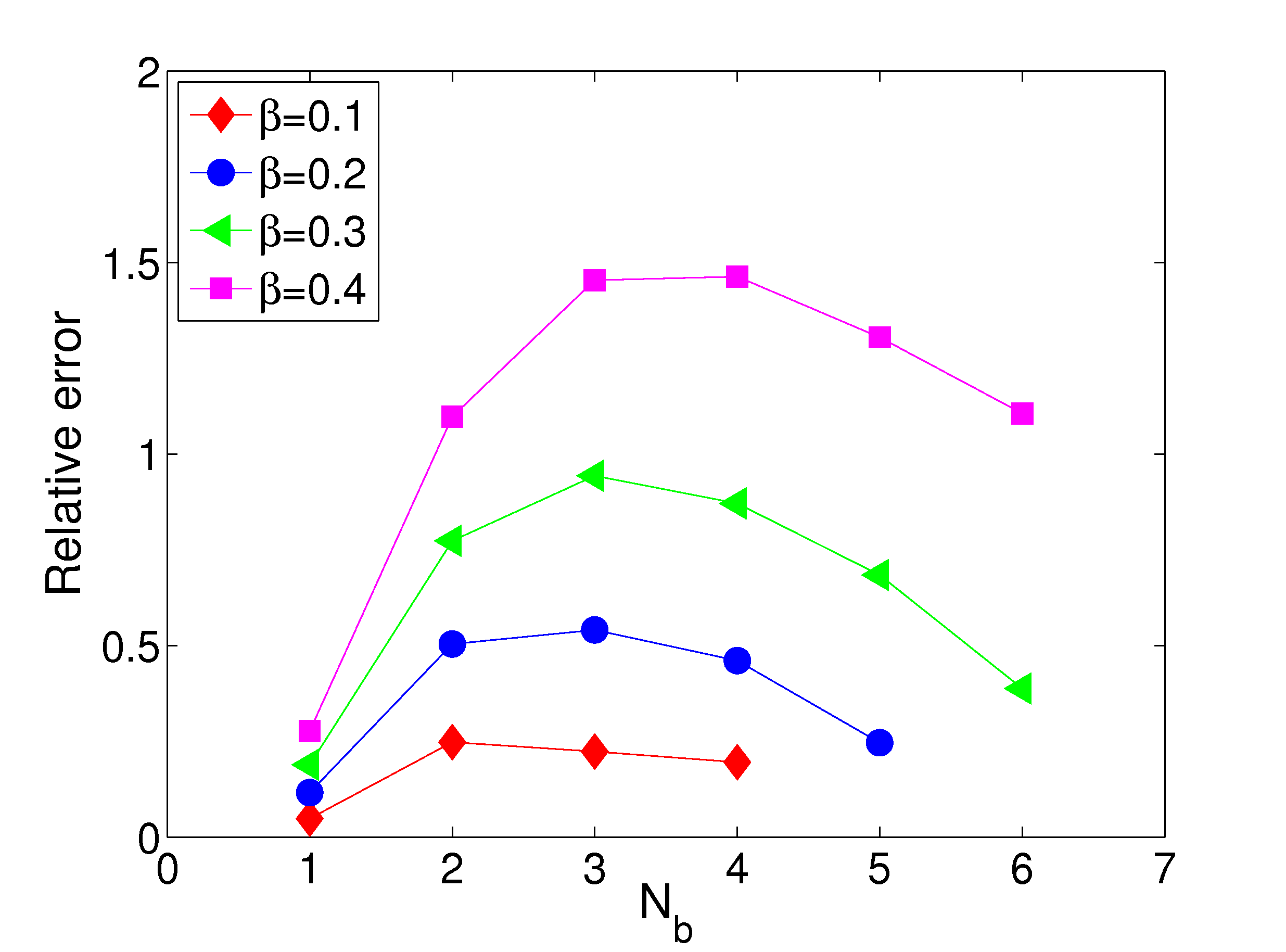}
\end{center}
\caption{
Relative error between the numerical result and
theoretical prediction for the decay rate of the
tail of $P(T)$.
Error is plotted against
$N_b$, for various 
$\beta$,  $L=100$ or $400$, $\alpha=0.6$.
Lines are guides to the eyes.
}
\label{fig:error}
\end{figure}

We focus on the jammed state, though
a similar analysis 
could be done in the FF phase.
If we assume as before that
a stationary state is established in the bottleneck, then
the probability for the
bottleneck to be empty should be $P_\emptyset=\beta^{N_b}$ at
each time step and,
if we neglect correlations between successive
time steps, the distribution $P(T)=(1-P_{\emptyset})^{T-1}
P_{\emptyset}$ follows.
For $N_b=1$, 
the stationary state is established very rapidly.
For larger $N_b$, non negligible 
corrections are present because there is a non vanishing relaxation time $\theta$ towards stationarity.
The accuracy of our prediction for $P(T)$ depends on the
ratio between the typical times $T$ and the relaxation time $\theta$.
The typical time $T$ is expected to increase
more rapidly with the bottleneck size $N_b$ than the relaxation
time $\theta$, making our prediction more accurate
for large $N_b$. Indeed, in Fig. \ref{fig:error} we observe that
the relative error between our prediction and simulation
results decreases again for large bottlenecks.
For all $N_b$, the prediction works better for small $\beta$,
and the difference between theory and numerics
increases when the MC phase is approached. 
Indeed, when $\beta$ increases, $P_\emptyset$ increases,
and the typical $T$ value decreases. It can then
become smaller than the relaxation time $\theta$
 and a description of transients
becomes necessary. Besides, the queues may not
have enough time to form and thus do not overfeed
the bottleneck.

Another feature visible in Fig. \ref{fig:distT} is that
there is an important contribution
to $P(T)$ from events during which only a single or very
few particles pass through the bottleneck. 
When these events occur, the system explores only very special
configurations, which are not representative of the whole
stationary distribution. Thus the system is still in a transient
state when the next reversal occurs.

As a conclusion the dynamics for the reversal of the flux actually involves
two quite different mechanisms (one based on stationarity
and the other involving transients), 
and exhibits some kind
of intermittency, with long periods of one-directional flows
alternating with rapid switches.
In order to have a full description of the dynamics
of the flux reversal, not only the transient nature of the
incoming flows in the bottleneck should be taken into
account, but also the correlations
between successive switches.

\section{Conclusions}

We have introduced a new model for bidirectional 
transport with a bottleneck.
We have shown that theoretical considerations,
based on the stationarity hypothesis for the
flow inside the
bottleneck, allow to predict with a good accuracy
the phase diagram,
and the values of the currents in the free
flow phase,  in the jammed phase,
and in the MC phase for long bottlenecks.
For short bottlenecks, transients
dominate the behavior of the system, and
as a consequence
large values of the current can be observed,
in spite of
the cost of having to empty the bottleneck at each
reversal of the flux.
These transients could be studied by exploiting the 
exact results obtained for a single TASEP with a step initial condition
\cite{asepvskpz,derrida_g09,tracy_w09}.

The jammed phase turns out to have a striped
structure. In spite of the
similarity with \cite{appert_s01}, here
the striped phase can be sustained in the bulk
without any modification of the standard TASEP bulk rules.
It should be noted that this striped phase results
from a self-regulated dynamics for the flux reversal
in the bottleneck.
Though an assumption of constant reversal periods
succeeds in predicting the striped phase density,
we find that actually the structure of
the distribution of the bottleneck occupation times $P(T)$
is more complex and emerges
from two different mechanism. 
The tail of the distribution can be explained
through our phenomenological approach assuming
a 1-lane stationary state in the bottleneck.
However, this prediction is valid only 
for small $\beta$, and in any case cannot explain 
the shape of the distribution around its maximum. 
A large part of the distribution is due to
non-typical
events where only a few particles go through the
bottleneck.
A complete understanding of $P(T)$ should
involve the study of non-stationary
distributions and correlations between flux reversals.

To conclude, the main feature that puts this
new model apart from others in the literature
is that 
fluctuations
localized in the bottleneck can have a macroscopic
effect on the whole system.
It provides a sensitive test for different theoretical approaches
and can be easily tested numerically.  
While a phenomenological approach assuming a stationary
state in the bottleneck gives surprisingly good predictions
for the free flow and part of the jammed phase, some other
observations trigger much more complex questions involving
non-stationary and correlated behaviors.
While we concentrated on 
identical lanes without directional bias, one can, of course, consider
asymmetries either in particles species or capacities
of the lanes which are of interest for different applications.

\begin{acknowledgments}
{
This work was supported by the French Research National
Agency (ANR) in the frame of the \mbox{PEDIGREE} contract
\mbox{(ANR-08-SYSC-015-01)}.\
L.S. (resp.\ A.J.) acknowledges support from the RTRA Triangle de la physique (Project \mbox{2010~--~027T}) (resp.\ Project \mbox{2011~--~033T}).
We thank B.~Derrida for inspiring discussions.
}
\end{acknowledgments}

\bibliographystyle{unsrt}

\end{document}